\documentclass[twocolumn]{revtex4}

\usepackage{amsmath}

\newcommand{\be}{\begin{eqnarray}}
\newcommand{\ee}{\end{eqnarray}}
\newcommand{\ket}[1]{\ensuremath{\left| {#1} \right>}}
\newcommand{\bra}[1]{\ensuremath{\left< {#1} \right|}}

\newcommand{\sx}{\ensuremath{\hat{\sigma}_x}\,}

\newcommand{\sz}{\ensuremath{\hat{\sigma}_z}\,}

\usepackage{latexsym}
\usepackage{graphicx}
\usepackage{amssymb}
\usepackage{multirow}
\usepackage{textcomp}

\begin{document}

\title{Time-dependent Hamiltonian estimation for Doppler velocimetry of trapped ions}

\author{L.~E.~de~Clercq, R.~Oswald, C.~Fl{\"u}hmann, B.~Keitch, D.~Kienzler,  H.-Y.~Lo, M. Marinelli,  D.~Nadlinger, V.~Negnevitsky,    J.~P.~Home}

\affiliation{Institute for Quantum Electronics, ETH Z\"urich, Otto-Stern-Weg 1, 8093 Z\"urich, Switzerland}


\pacs{pacs}
\maketitle

\textbf{The time evolution of a closed quantum system is connected to its Hamiltonian through Schr{\"o}dinger's equation. The ability to estimate the Hamiltonian is critical to our understanding of quantum systems, and allows optimization of control. Though spectroscopic methods allow time-independent Hamiltonians to be recovered, for  time-dependent Hamiltonians this task is more challenging \cite{14Zhang,03Mitra,10Rabitz,13FemtoChemRey,devitt2006scheme,11Shabani}. Here, using a single trapped ion, we experimentally demonstrate a method for estimating a time-dependent Hamiltonian of a single qubit. The method involves measuring the time evolution of the qubit in a fixed basis as a function of a time-independent offset term added to the Hamiltonian. In our system the initially unknown Hamiltonian arises from transporting an ion through a static, near-resonant laser beam \cite{07Leibfried}. Hamiltonian estimation allows us to estimate the spatial dependence of the laser beam intensity and the ion's velocity as a function of time. This work is of direct value in optimizing transport operations and transport-based gates in scalable trapped ion quantum information processing \cite{98Wineland2, 12Bowler, 12Walther}, while the estimation technique is general enough that it can be applied to other quantum systems, aiding the pursuit of high operational fidelities in quantum control \cite{14Martinis, 13Schutjens}.}

Estimation of the underlying dynamics which drive the evolution of systems is a key problem in many areas of physics and engineering. This knowledge allows control inputs to be designed which account for imperfections in the physical implementation.  For closed quantum systems, the time dependence of a system is driven by the Hamiltonian through Schr{\"o}dinger's equation. If the Hamiltonian is static in time, a wide range of techniques have been proposed for recovering the Hamiltonian \cite{05Cole,03Mitra,10Rabitz, 14Zhang}, which have been applied to a variety of systems including chemical processes \cite{13FemtoChemRey} and quantum dots \cite{devitt2006scheme,11Shabani}. These methods often involve estimation of the eigenvectors and eigenvalues of the Hamiltonian via spectroscopy, or through pulse-probe techniques for which a Fourier transform of the time-evolution gives information about the spectrum. However these methods are not directly applicable to time-dependent Hamiltonians. Such Hamiltonians are becoming of increasingly important as quantum engineering pursues a combination of high operational fidelities and speed, often involving fast variation of control fields which are particularly susceptible to distortion before reaching the quantum device \cite{12Zhao,04Blais,13Devoret,12Bowler,12Walther,ball2015walsh}.

In this Letter, we propose and demonstrate a method for reconstructing a general time-dependent Hamiltonian with two non-commuting terms which drives the evolution of a single qubit. The method works with any single qubit Hamiltonian $\hat{H} = \sum_i f_i(t) \hat{\sigma}_i$, where the $f_i(t)$ are arbitrary time-dependent functions and $\hat{\sigma}_i$ are the Pauli operators. In our experiments, a  Hamiltonian with two non-commuting time-dependent terms arises when we try to perform quantum logic gates by transporting an ion through a static laser beam \cite{07Leibfried, 15deClercq}. In this case, the Hamiltonian describing the interaction between the ion and the laser can be written in an appropriate rotating frame as
\be
\label{eq:Hamiltonian}
\hat{H}_I(t) = \frac{\hbar}{2}\left( - \Omega(t) \sx +  \delta(t) \sz\right)
\ee
which includes a time-varying Rabi frequency $\Omega(t)$, and an effective detuning $\delta(t)$ which is related to the first-order Doppler shift of the laser in the rest frame of the moving ion (see \cite{Methods2} for details). For a Hamiltonian of this type with unspecified time-dependent coefficients, no analytical solution to Schr{\"o}dinger's equation exists \cite{Barnes12,Barnes13}. In order to reconstruct the Hamiltonian we make use of two additional features of our experiment. The first is that we can switch off the Hamiltonian at time $t_{\rm off}$ on a timescale which is fast compared to the evolution of the qubit. Secondly we are able to offset one of the terms in the Hamiltonian, in our case by adding a static detuning term $\hat{H}_{\rm s} = \hbar\delta_L \sz/2$ such that the total Hamiltonian is $\hat{H}_I(t) + \hat{H}_s$. We then measure the expectation value of the qubit in the $\hat{\sigma}_z$ basis as a function of $\delta_L$ and $t_{\rm off}$. Repeating the experiment with identical settings many times, we obtain an estimate of the expectation value which we denote as $\langle \hat{\sigma}_z^{\rm meas} \left( t_{\rm off},\delta_L \right) \rangle$.

Hamiltonian extraction involves theoretically generating the qubit populations $\langle \hat{\sigma}_z^{\rm sim} \left( t_{\rm off},\delta_L \right) \rangle$, and attempting to find the Hamiltonian for which this most closely matches the data. In order to provide a simple parameterization, we represent $\delta(t)$ and $\Omega(t)$ as a linear weighted combination of basis splines \cite{95Bartels,deBoorSplines}. $\langle \hat{\sigma}_z^{\rm sim} \left( t_{\rm off},\delta_L \right) \rangle$ is compared to the measured data using a weighted least-squares cost function, which we optimize with respect to the weights of the basis-splines used to parameterize $\delta(t)$ and $\Omega(t)$. Solving this optimization problem in general is hard because the cost function is subject to strong constraints imposed by quantum mechanics, producing a non-trivial relation between the weights and the spin populations \cite{Methods2}. We overcome this problem by making use of the inherent causality of the quantum-mechanical evolution, and by assuming that the parameters of the Hamiltonian vary smoothly. We call our technique ``Extending the Horizon Estimation'', in analogy to established methods in engineering \cite{95Muske} (a detailed description of our method can be found in \cite{Methods2}). Rather than optimizing over the whole data set at once, we build up the solution by initially fitting the data over a limited region of time $0<t_{\rm off}<T_0$. The solution obtained over this first region can be extrapolated over a larger time span  $0<t<T_1$ where $T_1 = T_0 + \tau$, which we use as a starting point to find an optimal solution for this extended region. This procedure is iterated until $T_{n_{\rm max}} = {\rm max}(t_{\rm off})$. The method allows us to choose a reduced number of basis spline functions to represent $\delta(t)$ and $\Omega(t)$, and also reduces the amount of data considered in the early stages of the fit, when the least is known about the parameters. This facilitates the use of non-linear minimization routines, which are based on local linearization of the problem and converge faster near the optimum. More details regarding the optimization routine can be found in \cite{Methods2}.

 In the experimental work, we demonstrate reconstruction of the spin Hamiltonian for an ion transported through a near-resonant laser beam. Our qubit is encoded in the electronic states of a trapped calcium ion, which is defined by $\ket{0}\equiv\ket{^2S_{1/2}, M_J = 1/2}$ and $\ket{1}\equiv\ket{^2D_{5/2}, M_J = 3/2}$. This transition is well resolved from all other transitions, and has an optical frequency $\omega_0/(2 \pi) \simeq 411.0420$~THz. The laser beam points at 45~degrees to the transport axis, and has an approximately Gaussian spatial intensity distribution. The time-dependent velocity $\dot{z}(t)$ of the ion is controlled by adiabatic translation of the potential well in which the ion is trapped. This is implemented by applying time-varying potentials to multiple electrodes of a segmented ion trap, which are generated using a multi-channel arbitrary waveform generator, each output of which is connected to a pair of electrodes via a passive third order low-pass Butterworth filter. The result is that the ion experiences a time-varying Rabi frequency $\Omega(t)$ and a laser phase which varies with time as $\Phi(t) = \phi(z(t)) - \omega_L t$, where $\phi(z(t)) = k_z(z(t))z(t)$ with $k_z(z(t))$ the laser wavevector projected onto the transport axis at position $z(t)$ and $\omega_L$ the laser frequency. The spatial variation of $k_z(z(t))$ accounts for the curvature of the wavefronts of the Gaussian laser beam. In order to create a Hamiltonian of the form of equation \ref{eq:Hamiltonian}, we work with the differential of the phase, which gives a detuning $\delta(t) = \delta_L - \dot{\phi}  =  \left(k_z'(z) z + k_z(z)\right) \dot{z}$ with $\delta_L=\omega_L - \omega_0$ the laser detuning from resonance. For planar wavefronts $k_z'(z) = 0$, and $\delta(t)$ corresponds to the familiar expression for the first-order Doppler shift (see \cite{Methods2} for details).

\begin{figure}
	\includegraphics[width = 1\columnwidth]{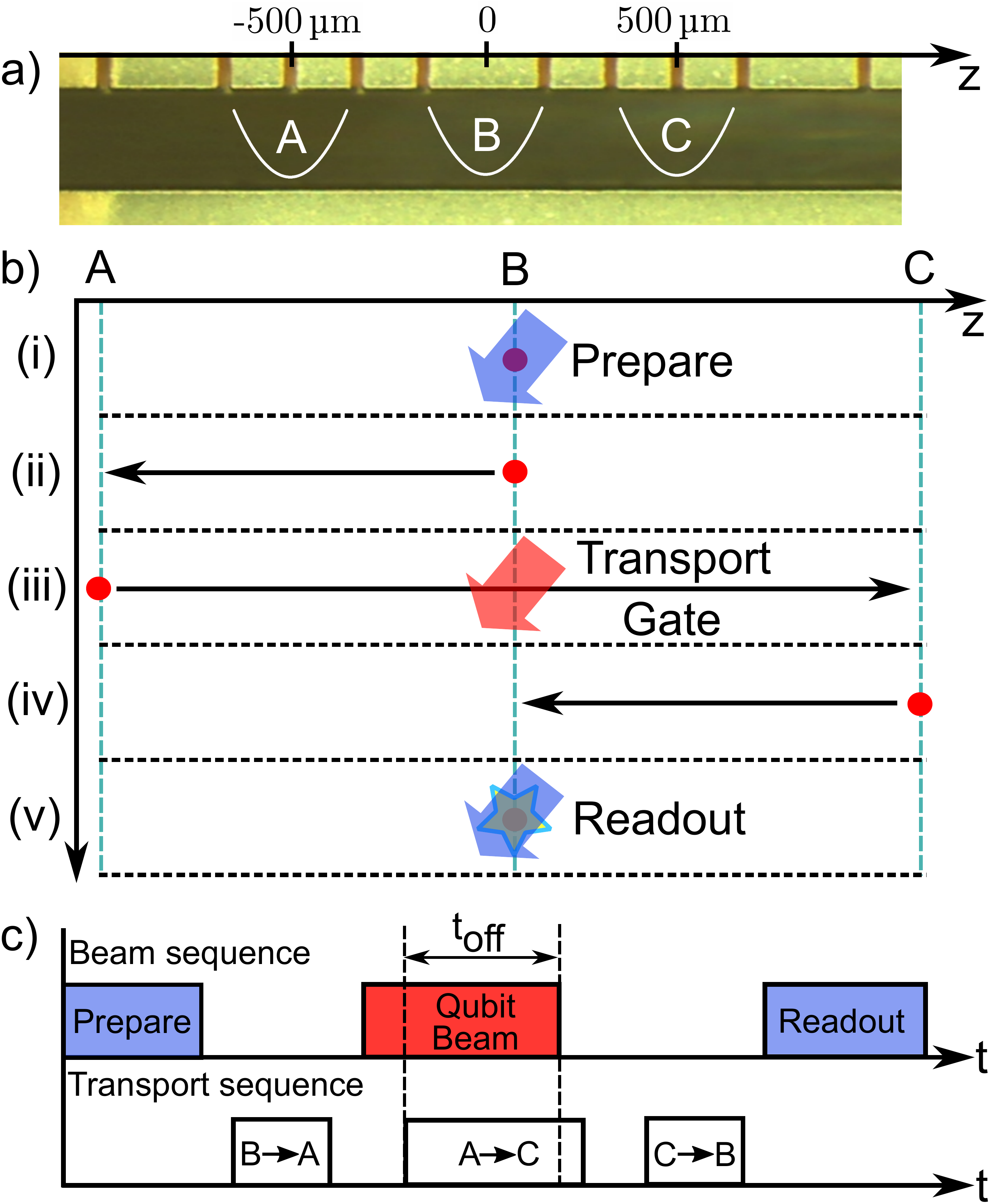}
	\caption{\textbf{Experimental sequence and timing:} a) The experiment is carried out in three zones of the trap indicated by A, B and C. b) The experimental sequence involves steps (i) through (v). Preparation and readout are carried out on the static ion in zone B. The qubit evolves while the ion is transported  from zone A to zone C, via the laser beam in zone B. c) Experimental sequence showing the timing of applied laser beams and ion transport, including shutting off the laser beam during transport.}
	\label{fig:exp_seq_and_timing}
\end{figure}

The experimental sequence is depicted in figure \ref{fig:exp_seq_and_timing}. We start by cooling all motional modes of the ion to $\bar{n} < 3$ using a combination of Doppler and electromagnetically-induced-transparency cooling \cite{00RoosEIT}, and then initialize the internal state by optical pumping into $\ket{0}$. The ion is then transported to zone A, and the laser beam used to implement the Hamiltonian is turned on  in zone B. The ion is then transported through this laser beam to zone C. During the passage through the laser beam, we rapidly turn the beam off at time $t_{\rm off}$ and thus stop the qubit dynamics. The ion is then returned to the central zone B in order to perform state readout, which measures the qubit in the computational basis (for more details see \cite{Methods2}). The additional Hamiltonian $\hat{H}_{\rm s}$ is implemented by offsetting the laser frequency used in the experiment by a detuning $\delta_L$. For each setting of $t_{\rm off}$ and $\delta_L$ the experiment is repeated 100 times, allowing us to obtain an estimate for the qubit populations $\langle \sz(t_{\rm off},\delta_L)\rangle$.

\begin{figure*}
	\includegraphics[width = 1.95\columnwidth]{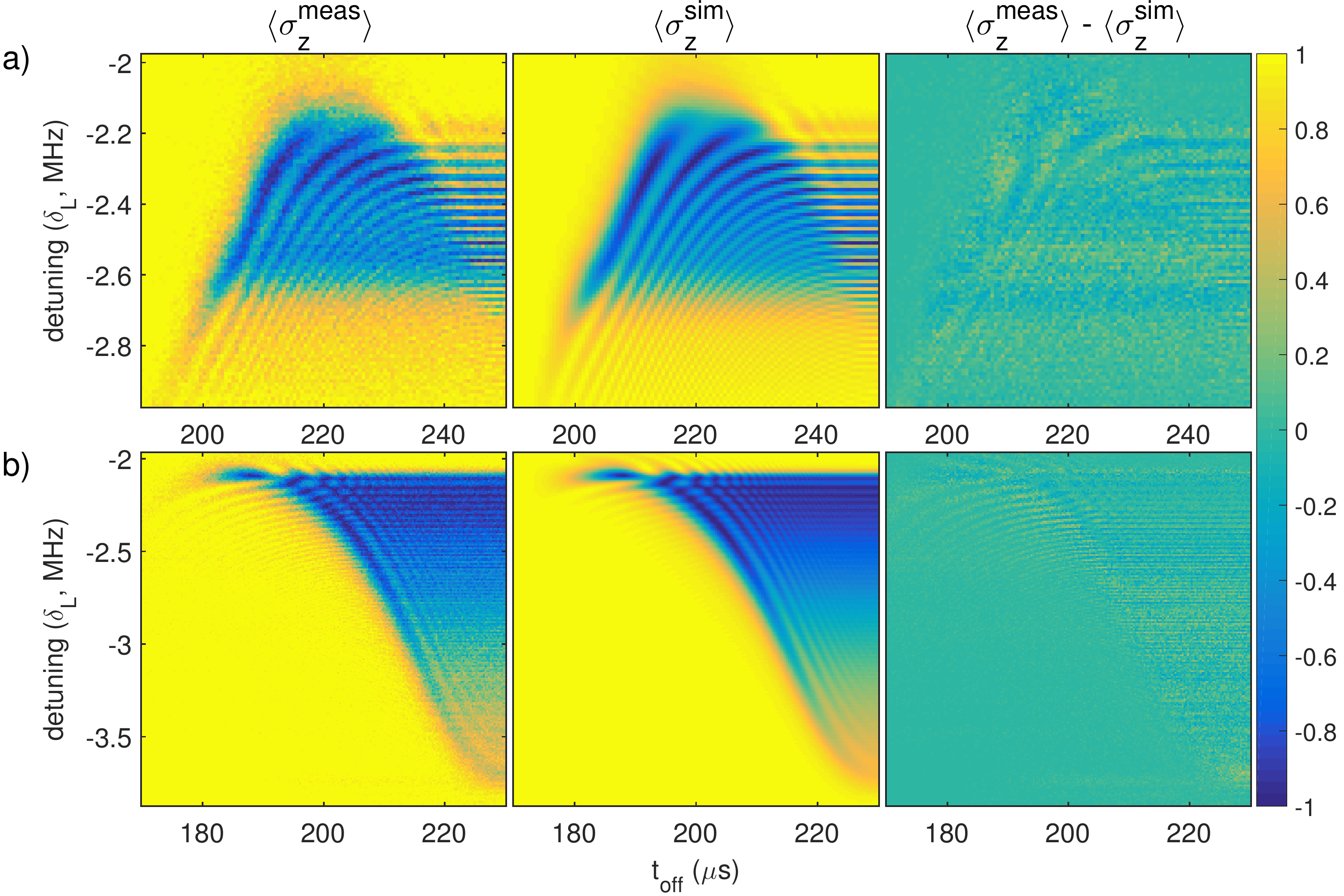}
	\caption{\textbf{Measured data, estimation and residuals:} Spin population as a function of detuning and switch-off time of the laser beam. a) is for a laser beam centered in zone B, while for b) the beam was displaced towards zone C by 64~$\mu$m. From left to right are plotted the experimental data, the populations generated from the best fit Hamiltonian, and the residuals. Each data point results from 100 repetitions of the experimental sequence. The data in a) consist of an array of $100\times101$ experimental settings, while that shown in b) consists of an array of $201\times201$ settings. This leads to smaller error bars in the reconstructed Hamiltonian for the latter. For the Hamiltonian estimation the data was weighted according to quantum projection noise.}
	\label{fig:Exp_Sim_Res}
\end{figure*}

\begin{figure}
	\includegraphics[width = 1\columnwidth]{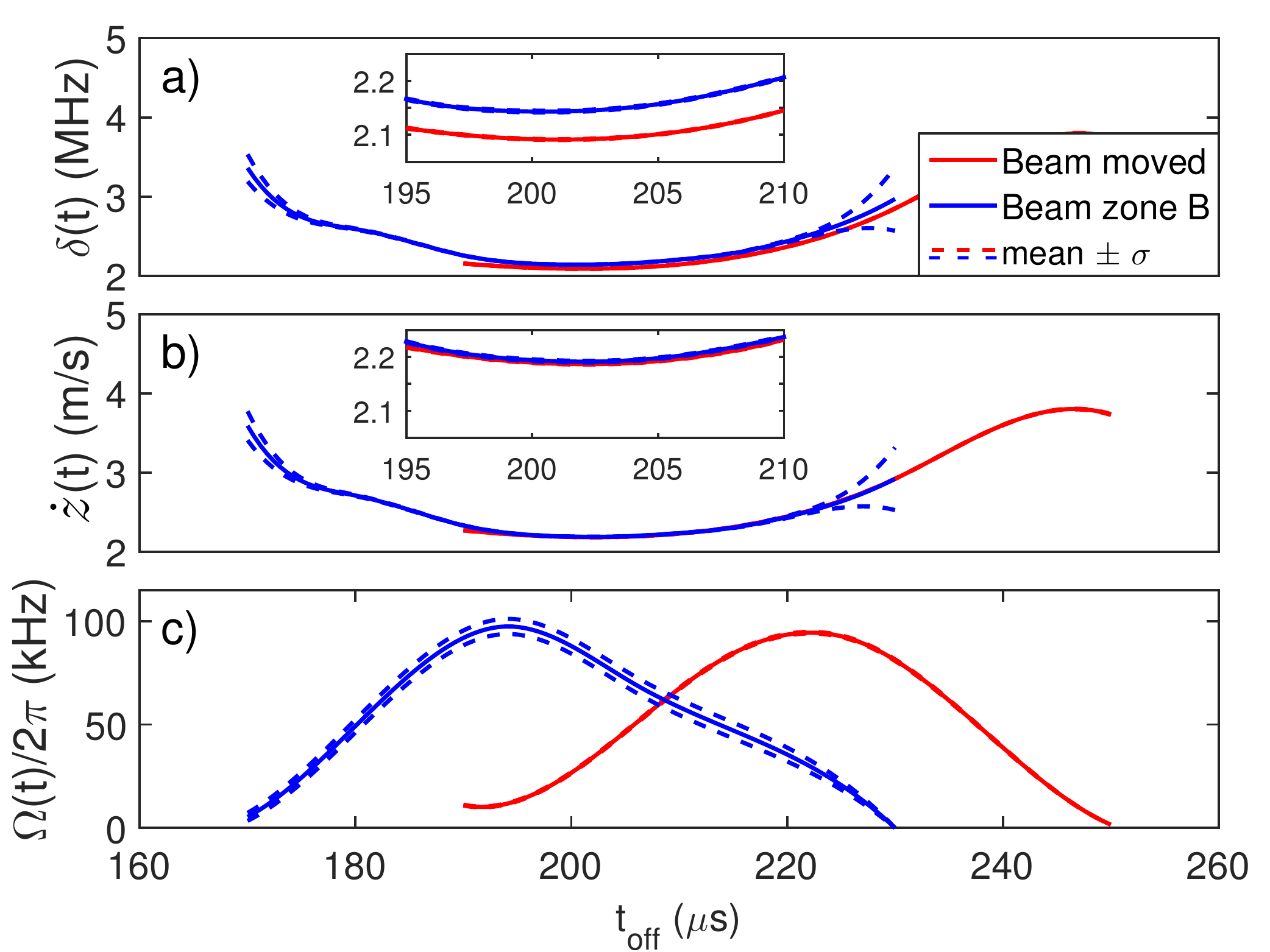}
	\caption{\textbf{Estimates of time-dependent co-efficients:} a) The effective detuning $\delta(t)$ and c) Rabi frequency $\Omega(t)$ obtained from the two data sets, along with dashed lines indicating the standard error on the mean of these estimates obtained from resampling. For a), the inset shows a close up of the estimated $\delta(t)$ in the regions where the estimates overlap, showing that these do not give the same value. b) The estimated velocity $\dot{z}(t)$ of the ion obtained after applying wavefront correction. The inset shows that this can produce consistent results.}
	\label{fig:before_after_beams}
\end{figure}


\begin{figure}
	\includegraphics[width = 1\columnwidth]{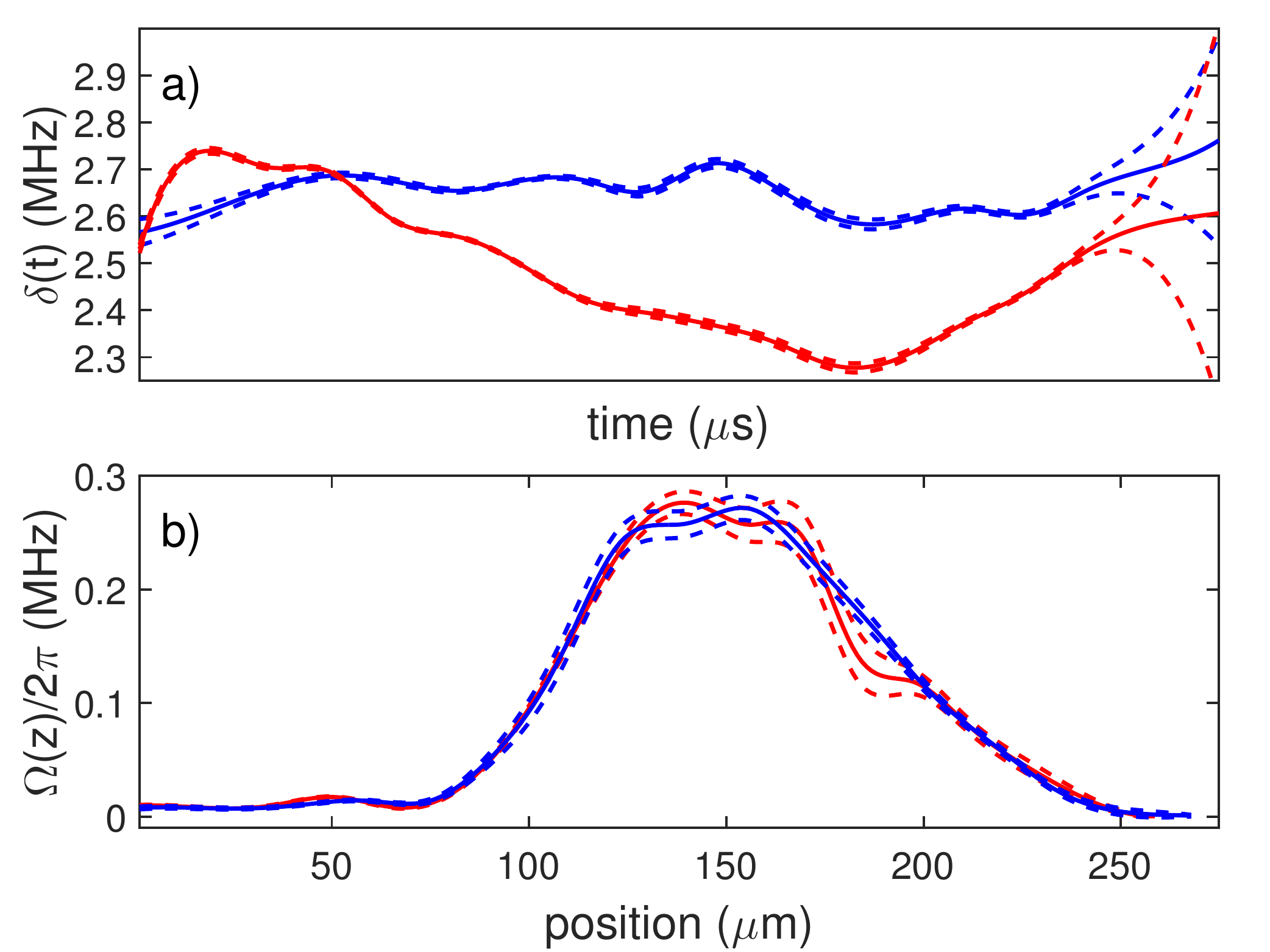}
	\caption{\textbf{Spatial Rabi frequency:} a) The estimated $\delta(t)$ obtained from the second pair of data sets (Figure \ref{fig:seconddata} in \cite{Methods2}).  b) The estimated Rabi frequency $\Omega(t)$ for the same two data sets.}
	\label{fig:spatial_rabi_freqs}
\end{figure}

Experimental data is shown  in figure \ref{fig:Exp_Sim_Res} for two different beam positions, alongside the results of fitting performed using our iterative method. The beam positions used for each data set differ by around 64~\textmu m, but the transport waveform used was identical. The reconstructed velocities should therefore agree in the region where the data overlap. It can be seen from the residuals that the estimation is able to find a Hamiltonian which results in a close match to the data. In order to get an estimate of the relevant error bars for our reconstruction, we have performed non-parametric resampling with replacement, optimizing for the solution using the same set of B-spline functions as was used for the experimental data to provide a new estimate for the Hamiltonian. This is repeated for a large number of samples, resulting in a distribution for the estimated values of $\delta(t)$ and $\Omega(t)$ from which we extract statistical properties such as the standard error. The error bounds shown in figures \ref{fig:before_after_beams} correspond to the standard error on the mean obtained from these distributions (see \cite{Methods2} for further details).

The estimated coefficients of the Hamiltonian extracted from the two data sets are shown in figure \ref{fig:before_after_beams}a). It can be seen that the values of $\delta(t)$ for the two different beam positions differ for the region where the reconstructions overlap. We think that this effect arises from the non-planar wavefronts of the laser beam. Inverting the expression for $\delta(t)$ to obtain the velocity of the ion, we find $\dot{z}(t) = \delta(t)/(k_z'(z)z + k_z(z))$. Using this correction, we find that the two velocity profiles agree  if we assume that the ion passes through the center of the beam at a distance of 2.27~mm before the minimum beam waist, a value which is consistent with experimental uncertainties due to beam propagation and possible mis-positioning of the ion trap with respect to the fixed final focusing lens. The velocity estimates taking account of this effect are shown in figure \ref{fig:before_after_beams}b).

Figure \ref{fig:spatial_rabi_freqs} shows the results of a reconstruction for a second pair of data sets taken using two different velocity profiles but with a common beam position. The resolution in both time and detuning were lower in this case than for the data shown in figure \ref{fig:Exp_Sim_Res} (see \cite{Methods2} for the data). We observe that the estimated Rabi frequency profiles agree to within the error bars of the reconstruction. One interesting feature of this plot is that the error bars produced from the resampled data sets are notably higher at the peak than on the sides of the beam. We think that this happens because the sampling time of the data is 0.5~$\mu$s, which is not high enough to accurately resolve the fast population dynamics resulting from the high Rabi frequency (the Nyquist frequency is 1~MHz). In order to optimize the efficiency of our method, it would be advantageous to run the reconstruction method in parallel with data taking, thus allowing updating of the sampling time and frequency resolution of points based on the current estimates of parameter values.

Our method for directly obtaining a non-commuting time-dependent Hamiltonian uses straightforward measurements of the qubit state in a fixed basis as a function of time and a controlled offset to the Hamiltonian. This simplicity means that the method should be applicable in a wide range of physical systems where such control is available, including many technologies considered for quantum computation \cite{devitt2006scheme,11Shabani,14Zhang,15Barends,15Bonato}. A process-tomography based approach would require that for every time step multiple input states be introduced, and a measurement made in multiple bases \cite{97NielsenChuang,97Poyatos,06Riebe}. An effective modulation of the measurement basis arises in our approach due to the additional detuning $\delta_L$. It is worth noting that tomography provides more information than our method: it makes no assumptions about the dynamics aside from that of a completely positive map while we require coherent dynamics. Extensions to our work are required in order to provide a rigorous estimation of the efficiency of the method in terms of the precision obtained for a given number of measurements, and to see whether a similar approach could be taken to non-unitary dynamics. Using this method on
considerably lower resolution data sets, we have recently been able to improve the control over the velocity, which will be necessary in order to realize multi-qubit transport gates in our current setup \cite{07Leibfried}.

We thank Lukas Gerster, Martin Sepiol and Karin Fisher for contributions to the experimental apparatus. We acknowledge support from the Swiss National Science Foundation under grant numbers $200021\_134776$ and $200020\_153430$, ETH Research grant under grant no. ETH-18 12-2, and from the National Centre of Competence in Research for Quantum Science and Technology (QSIT).\\

\textbf{Author Contributions:} Experimental data was taken by LdC, RO and MM using apparatus built up by all authors. Data analysis was performed by LdC and RO. The paper was written by JPH, LdC and RO, with input from all authors. The work was conceived by JPH and LdC.

\bibliographystyle{unsrt}
\bibliography{./myrefs_add}

\begin{thebibliography}{10}
\bibitem{Methods2}
Materials and methods are available online.
\end{thebibliography}

\clearpage
\newpage
\clearpage
\newpage
\section{Supplementary material}

\subsection{Derivation of Hamiltonian}
The interaction of a laser beam with frequency $\omega_L$ and  wave vector $\vec{k}(\vec{z}(t))$ with a two-level atom with resonant frequency $\omega_0$ and time-dependent position of the ion $\vec{z}(t)=(0,0,z(t))$ can be described in the Schr\"odinger picture by the Hamiltonian
\be
  \hat{H}_{S} = -\frac{\hbar \omega_0}{2} \sz - \hbar \Omega(z(t)) \cos \left(  \vec{k}(\vec{z}(t)) \cdot \vec{z}(t)- \omega_{L}t \right) \sx,
  \label{eq:Ham_transport_basic1}
\ee
where the Rabi frequency $\Omega(z(t))$ gives the interaction strength between the laser and the two atomic levels. We can define the laser phase at the position of the ion as $\Phi(t) = \phi(t)- \omega_{L}t$ with $\phi(t)= \vec{k}(\vec{z}(t)) \cdot \vec{z}(t) =k_z(z(t))z(t)$ and $k_z(z(t)) = |\vec{k}|\cos \left(\theta(t)\right)$ being the projection of the laser beam onto the $z$-axis along which the ion is transported. Here $\theta(t)$ is the angle between the wave-vector $\vec{k}(z(t))$ and the transport axis evaluated at position $z(t)$. Moving to a rotating frame using the unitary transformation $U = e^{-i\frac{\Phi(t)}{2}}$ and applying the  rotating wave approximation with respect to optical frequencies, we obtain
\be
\hat{H}_{I} = \frac{\hbar}{2} \left(- \Omega(t)\sx + \left(- \omega_0-\dot{\Phi}(t)\right)\sz \right).
\ee
Defining a static detuning $\delta_{L}=\omega_L - \omega_0$ we obtain
\be
\hat{H}_{I} = \frac{\hbar}{2} \left( -\Omega(t)\sx + \left(\delta_L-\dot{\phi}(t)\right)\sz \right).
\ee
with
\be
\delta(t) = \delta_L-\dot{\phi}(t)
\label{eq:delta_corr_eqn}
\ee
which is the expression used in the main text.

\subsection{Wavefront correction}
\begin{figure}
	\includegraphics[width = 1\columnwidth]{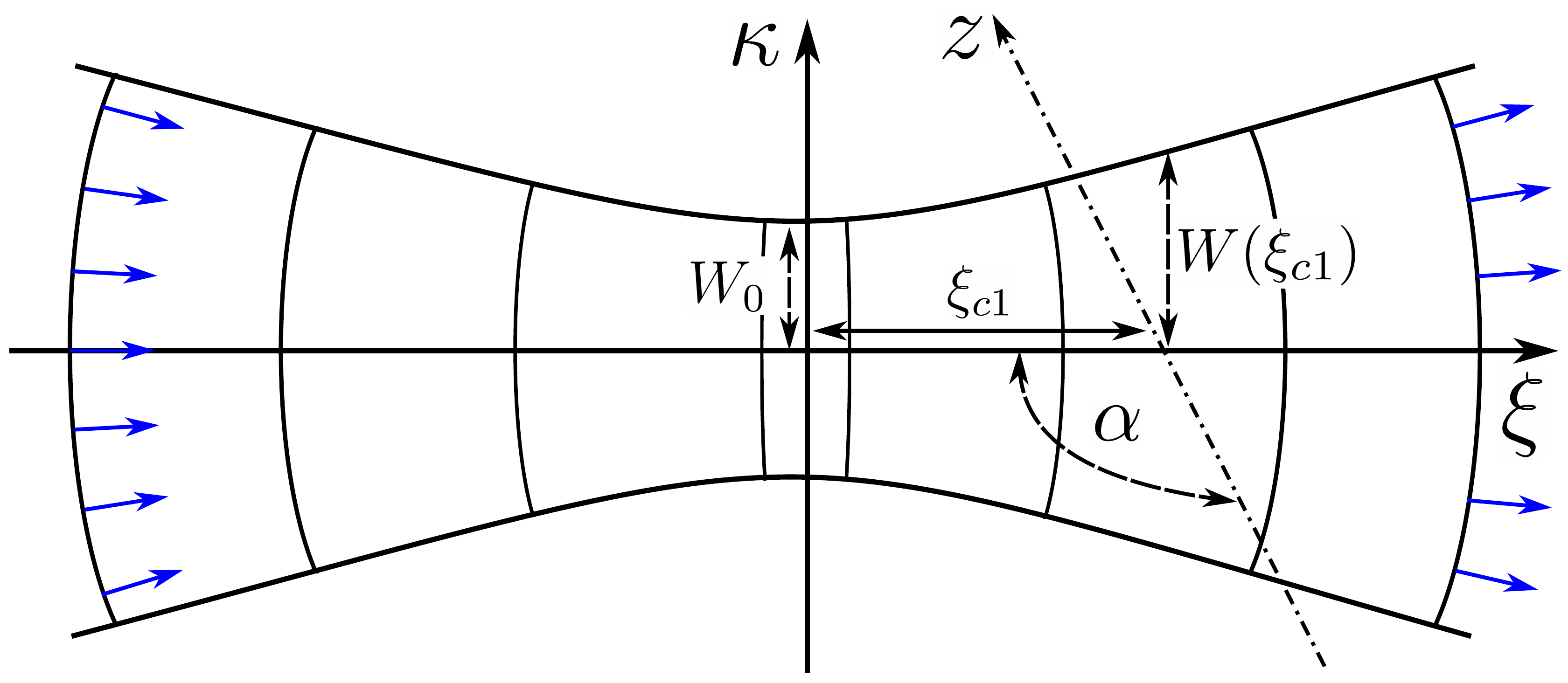}
	\caption{\textbf{Beam and ion transport:} The beam propagation direction lies along the $\xi$-axis and the ion is transported along the z-axis lying the $\kappa\xi$-plane as indicated. Normalized vectors representing $\vec{e}_l (\kappa,\xi)$ lying perpendicular to the wavefronts are indicated by the blue arrows.  }
	\label{fig:Gaussian_beam_depiction}
\end{figure}
For plane waves we find that $\dot{\phi}(t) = \vec{k}\cdot \vec{v}(t)$  which is the well-known expression for the first-order Doppler shift. For transport through a  real Gaussian beam, the wave-vector direction changes with position. Taking this into account, the derivative of  $\phi(t)$ becomes
\be
\dot{\phi}(t) = \left[k_z'(z(t))z(t) + k_z(z(t))\right] \dot{z}(t)
\ee
where $k_z'=dk_z/dz$ and $\dot{z}(t)$ is the component of the ion's velocity which lies along the $z$-axis. We extract $\delta(t)$ using our Hamiltonian estimation procedure, thus to obtain the velocity of the ion we use
\be
\dot{z}(t) = \frac{-\delta(t)+\delta_{L}}{k_z'(z(t))z(t) + k_z(z(t))} \ .
\ee
As the ion moves through the beam it experiences the same magnitude of the wave vector $|\vec{k}|=2\pi/\lambda$, but the angle $\theta$ between the ion's direction and the wave vector changes. Written as a function of this angle, the velocity becomes
\be
\dot{z}(t) = \frac{-\delta(t)+\delta_{L}}{ -|\vec{k}| \sin\left(\theta(z(t))\right) \theta'(z(t)) z(t) + |\vec{k}| \cos(\theta(z(t))) }
\label{eq:correction_eqn}
\ee
where $\theta'(z(t))=d\theta(z(t))/dz(t)$. We parameterize our Gaussian beam according to figure \ref{fig:Gaussian_beam_depiction}. The phase is given as a function of both the position along the beam axis $\xi$ and the perpendicular distance from this axis $\kappa$ by \cite{07Saleh}
\be
\varphi (\kappa,\xi) = |\vec{k}|\xi  -  \zeta(\xi) + \frac{|\vec{k}|\kappa^2}{2R(\xi)} \ .
\label{eq:gauss_phase}
\ee
where the Gaussian beam parameters include the beam waist $W(\xi)$, the radius of curvature $R(\xi)$ the Rayleigh range $\xi_R$ and the Guoy phase shift $\zeta(\xi)$. These are given by the expressions
\be
\begin{split}
W(\xi) &= W_0 \sqrt{1 + \left( \frac{\xi}{\xi_R} \right)^2} \label{eq:beam_waist}\\
R(\xi) &= \xi \left( 1 + \left( \frac{\xi_R}{\xi} \right)^2 \right) \\
\zeta(\xi) &= \tan^{-1} \left(\frac{\xi}{\xi_R}\right) \\
\xi_R &= \frac{\pi W_0^2}{\lambda} \label{eq:rayleigh_length}\\
k &= \frac{2\pi}{\lambda}
\end{split}
\ee
where $W_0$ is the minimum beam waist and $\lambda$ the laser wavelength. The ion moves along the $z$-axis shown in figure \ref{fig:Gaussian_beam_depiction}. In the $\kappa\xi$-plane a unit vector $\vec{e}_l(\kappa,\xi)$ perpendicular to the wavefronts is given by
\be
\vec{e}_l (\kappa,\xi) = \frac{\nabla \varphi (\kappa,\xi) }{||\nabla \varphi (\kappa,\xi)||}
\label{eq:phase_x_y}
\ee
and the unit vector $\vec{e}_v$ pointing along the direction of transport is given by
\begin{align}
     \vec{e}_v =\begin{bmatrix}
     \cos(\alpha) \\
     \sin (\alpha) \\
     \end{bmatrix}
\end{align}

The angle $\theta(\xi)$ between the wave and position vector is then given by the dot product
\be
\theta(\kappa) = \cos^{-1} \left( \vec{e}_n \cdot \vec{e}_v \right).
\ee
which can be written in terms of the full set of parameters above as
\begin{widetext}
\begin{align}
\begin{split}
\theta(\kappa) &= \cos^{-1} \left( \gamma_1 + \gamma_2 \right)\\
\gamma_1 &= \frac{\cos(\alpha)\left(-2\xi_R \left(\xi^2 + \xi_R^2\right) + k\kappa^2 \left(\xi_R^2 - \xi^2\right) + 2k \left(\xi^2 + \xi_R^2\right)^2\right)}{\eta(\kappa)} \\
\gamma_2 &= \frac{\sin(\alpha)2k\kappa \xi \left( \xi^2 + \xi_R^2 \right)}{\eta(\kappa)} \\
\eta(\kappa,\xi) &=\left(\xi^2 + \xi_R^2\right) \left[ 4 \left( \frac{k\kappa \xi}{\xi^2 + \xi_R^2} \right)^2 + \left(  -\frac{2\xi_R}{\xi^2 + \xi_R^2} + k \left(2 + \frac{\kappa^2 \left(\xi_R^2 - \xi^2\right)}{\left(\xi^2 + \xi_R^2\right)^2}\right) \right)^2\right] \\
\kappa(t) &= z(t)\sin(\alpha)
\label{eq:angle_k_v}
\end{split}
\end{align}
\end{widetext}
where in our experiments $\alpha = 3\pi/4$.

Using Eq. \ref{eq:correction_eqn} and \ref{eq:angle_k_v} we examined the value of $\xi_{\rm cl}$ required for the velocity to match for our two beam positions. We find that they agree for $\xi_{\rm cl} = -2.27$~mm, which is within the experimental uncertainties for our setup.
	
\subsection{Basis spline curves}
\begin{figure}
	\centering
	\includegraphics[width = 1\columnwidth]{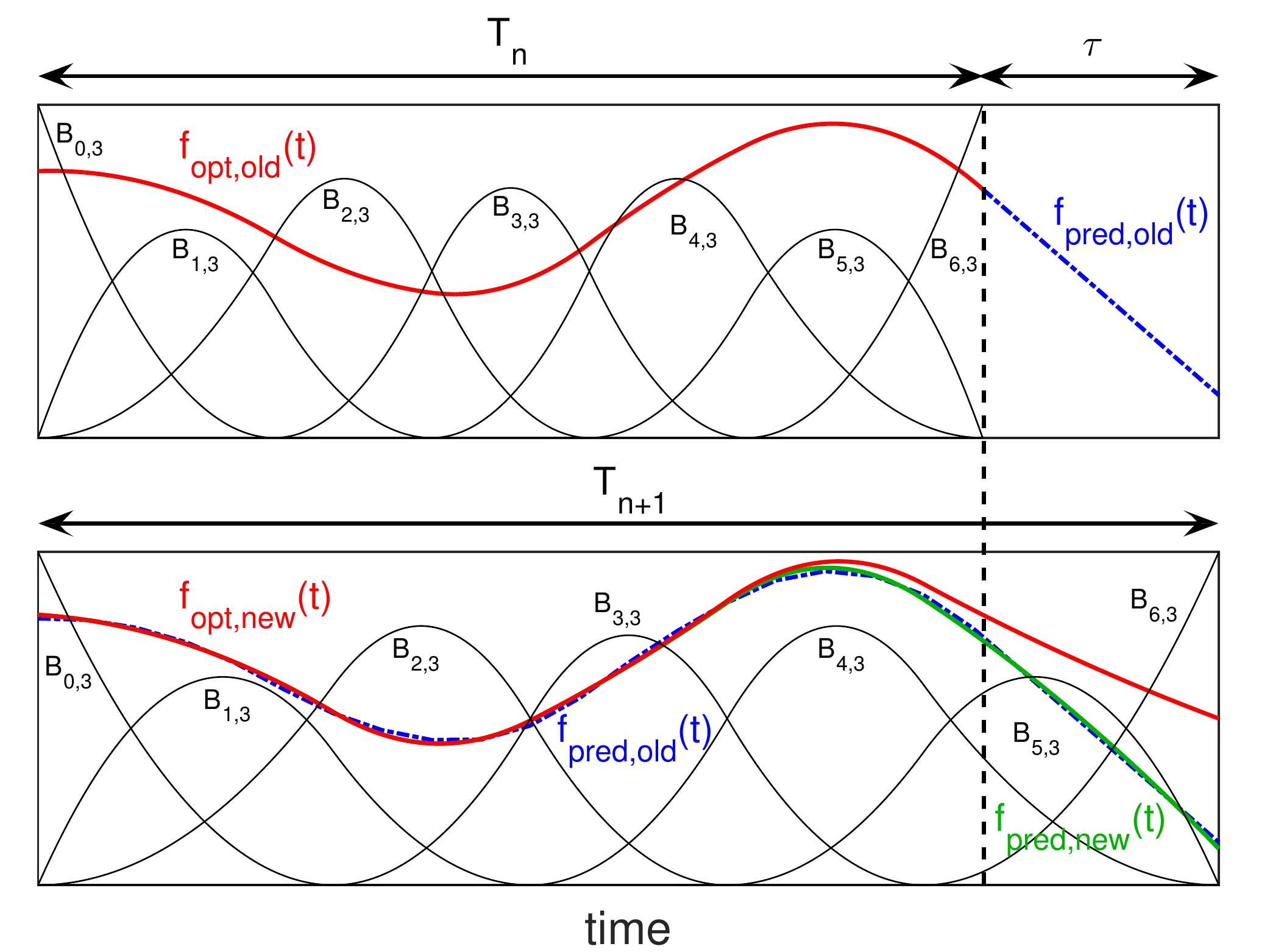}
	\caption{\textbf{Extending the Horizon Estimation:} The steps performed when extending the time horizon from $T_{\rm n}$ to $T_{\rm n+1}$ are illustrated. We first predict in the old basis, then move to the new basis, and finally optimize again. The figure also shows the basis splines $B_{i,k}(t)$.}
	\label{fig:splines_repr}
\end{figure}
One challenge in obtaining an estimate for the Hamiltonian is that we must optimize over continuous functions $\delta(t)$ and $\Omega(t)$. To address this, we represent $\delta(t)$ and $\Omega(t)$ with basis spline curves. Basis spline curves allow the construction of smooth functions using only a few parameters. This is achieved by introducing a set of polynomial Basis (B)-spline functions $B_{i,k}(t)$ of order $k$ \cite{deBoorSplines}. A smooth curve $S(t)$ can then be represented as a linear combination of these B-spline basis functions \cite{95Bartels}
\begin{align}
S(t) &= \sum\limits_{i=0}^{n} \alpha_{i} B_{i,k} (t).
\label{eq:spline_curve}
\end{align}
The B-splines $B_{i,k}(t)$ of order $k$ are recursively defined over the index $i$ over a set of points $\vec{K}=\{t_0,t_1,...,t_{n+k}\}$ which is referred to as the knot vector \cite{deBoorSplines}.
\begin{align}
\begin{split}
   B_{i,1}(t) &= \left\{
     \begin{array}{lr}
       1 & t_i \le t \le t_{i+1}\\
       0 & \text{otherwise}
     \end{array}
   \right. \\
   B_{i,k} (t) &=  \omega_{i,k} (t) B_{i,k-1} (t) + \left( 1-\omega_{i+1,k} (t)\right) B_{i+1,k-1} (t). \\
   \omega_{i,k} (t) &= \left\{
        \begin{array}{lr}
          \frac{t-t_i}{t_{i+k-1}-t_{i}} & \text{if  }t_i\ne t_{i+k-1}\\
          0 & \text{otherwise}
        \end{array}
    \right.
\end{split}
\label{eq:spline_rec}
\end{align}
Figure \ref{fig:splines_repr} gives a visualization of the B-splines $B_{i,k}(t)$ and a basis spline curve. The B-spline construction ensures that any linear combination of the B-splines is continuous and has $(k-2)$ continuous derivatives. The knot vector $\vec{K}$ determines how the basis functions are positioned within the interval $[t_0,t_{n+k}]$. We notice that for our Hamiltonian the spacing of the B-splines is not critical, which we think is due to the smoothness of the variations in our Hamiltonian parameters $\delta(t)$ and $\Omega(t)$. We therefore used the Matlab function \textit{spap2} to automatically choose a suitable knot vector and restricted ourselves to optimizing the coefficients $\alpha_{i}$. We collect all  coefficients $\alpha_{i}$ for $\delta(t)$ and $\Omega(t)$ and store them in a single vector $\vec{\alpha}$.

\subsection{Extending the Horizon Estimation}
The task of inferring the time-dependent Hamiltonian of the form (\ref{eq:Hamiltonian}) from the measured data can be cast into an optimization problem for which we use a reduced chi-squared cost function
\be
\label{eq:J}
J = \frac{1}{\nu} \sum_{t_{\rm off}} \sum_{\delta_L} \left[   \frac{  \langle \hat{\sigma}_z^{\rm meas} \left( t_{\rm off},\delta_L \right) \rangle - \langle \hat{\sigma}_z^{\rm sim} \left( t_{\rm off},\delta_L \right) \rangle     }{\sigma^{\rm meas}{\left( t_{\rm off},\delta_L \right)}} \right] ^2
\ee
where $\nu=N-n-1$ is the degrees of freedom with $N$ the number of data points and $n$ the number of fitting parameters, and $\sigma^{\rm meas}{\left( t_{\rm off},\delta_L \right)}$ is the standard error on the estimated $\langle \hat{\sigma}_z^{\rm meas} \left( t_{\rm off},\delta_L \right) \rangle$ which we obtain assuming quantum projection noise.  In our case the Hamiltonian $\hat H_I$ (Eq. \ref{eq:Hamiltonian}) with offset $\hat{H}_{\rm s} = \hbar\delta_L \sz/2$ is parametrized by $\Omega(t)$ and $\delta(t)$. We can thus write the problem as
\begin{equation}
\label{eq:opt}
 \min_{\delta(t), \, \Omega(t)}J(\delta(t),\Omega(t))
\end{equation}
subject to
\begin{eqnarray}
i \hbar \frac{\partial}{\partial t} \ket{\Psi(t,\delta_L)} &=& \left(\hat H_I(t) + \hat H_s\right)  \ket{\Psi(t,\delta_L)} , \nonumber \\
 \ket{\Psi(t=0,\delta_L)} &=& \ket{0} , \nonumber \\
 \langle \hat{\sigma}_z^{\rm sim} \left(t,\delta_L \right) \rangle &=& \bra{\Psi(t,\delta_L)} \sz \ket{\Psi(t,\delta_L)}
\end{eqnarray}
for all $\delta_L$.

This optimization problem is hard to efficiently solve in general, because it is nonlinear and non-convex due to the nature of Schr{\"o}dinger's equation and the use of projective measurements.  In order to overcome this challenge, we have implemented a method which we call ``Extending the Horizon Estimation" (EHE) in analogy to a well-established technique called ``Moving Horizon Estimation" (MHE) \cite{95Muske}.

The key idea is that because our measurement data arises from a causal evolution, we can also estimate the Hamiltonian in a causal way. We define a time span ranging from the initial time to some later time which we call the time horizon. Instead of optimizing $J$ over the complete time span at once, we first restrict ourselves to a small, initial time horizon reaching only up to the start of the qubit dynamics. Optimizing $J$ over this short time horizon requires fewer optimization parameters and is simpler than attempting to optimize over the full data set. Once we have solved this small sub-problem, we extend the time horizon and re-run the optimization, extrapolating the results of the initial time window into the extended window in order to provide good starting conditions for the subsequent optimization. This is greatly advantageous for the use of non-linear least squares optimization, which typically works by linearizing the problem and converges much faster near the optimum. The extension of the horizon is used repeatedly until the time window covers the full data set.

Conceptually EHE is very similar to MHE. The main difference is that in MHE the time span has a fixed length and thus its origin gets shifted forwards in time along with the horizon. In EHE the origin stays fixed at the expense of having to increase the time span under consideration. MHE avoids this by introducing a so-called arrival cost to approximate the previous costs incurred before the start of the time span. This keeps the computational burden fixed over time, which is very important as MHE is usually used to estimate the state of a system in real-time, often on severely constrained embedded platforms. Since neither constraint applies to our problem, we decided to extend the horizon rather than finding an approximate arrival cost. This is advantageous since finding the arrival cost in the general case is still an open problem. Due to the similarity between MHE and EHE, we anticipate future improvements by adapting techniques used in MHE to EHE.

Next, we present a more detailed algorithmic summary of our implementation of the method outlined above.
\begin{enumerate}
\item \textbf{Searching for a starting point.}
Here we reconstruct the Hamiltonian for a first, minimal time horizon such that we can then use this as a starting point to iteratively extend the horizon as described in step 2.
\begin{enumerate}
\item Choose an initial time horizon such that it contains the region where the first discernible qubit dynamics occur.
\item Cut down the number of fitting parameters as much as possible, e.g. by using few Basis splines of low order. This amounts to choosing empirically a low number of basis splines (and thus the length of $\vec{\alpha}_0$) which might represent $\delta(t)$ and $\Omega(t)$ over the given region.
\item Use a nonlinear least squares fitting routine to minimize $J$ by varying the parameters $\vec{\alpha}_0$. In the case that the initial fit is not good or no minimum is found, try new initial conditions, change the number of B-spline functions, or manually adjust the function using prior knowledge of the physical system under consideration.
\end{enumerate}

This procedure is used to provide a starting point for the optimization over the initially chosen window, which is typically performed with a higher order set of B-splines. From this starting point, we iteratively extend the fitting method to the full data set as follows.
\item \textbf{Extend the horizon}
This step is repeated until the whole time horizon is covered. It consists of the following sequence, which is illustrated in figure \ref{fig:splines_repr}.
\begin{enumerate}
\item Extend the time horizon by $\tau$ from $T_{\rm n}$ to $T_{\rm n+1} = T_{\rm n} + \tau$.
\item Extrapolate $f_{\rm opt,old}(t)$ within $\tau$, e.g. using \textit{fnxtr} in Matlab.
\item Adapt the Basis splines to the new time horizon $T_{\rm n+1}$ and represent $f_{\rm pred,old}(t)$ in the new basis, giving $f_{\rm pred,new}(t)$. In Matlab one can use \textit{spap2} to do this.
\item Use $f_{\rm pred,new}(t)$ as the initial guess for a weighted nonlinear least squares fit over the extended time span up to $T_{\rm n+1}$.
\item Judge the results of the fit based on its reduced chi-squared value $\chi_{red}^{2}$. If it is below a specified bound, continue with an additional iteration of steps a)-d), repeating until the full region of the data is covered. Otherwise, try the following fall-back procedures:
\begin{enumerate}
\item Reduce $\tau$, the time by which the time horizon is extended, and try again 
\item Increase the number of Basis splines and try again
\item Try again using a different starting point.
\end{enumerate}
If all of those fail, we have to resort to increasing the bound on $\chi_{\rm red}^{2}$.
\end{enumerate}

\item \textbf{Post-processing}.
The following steps are optional and were performed manually in cases where we desired to improve the fit, or examine its behaviour.
\begin{enumerate}
\item The optimization over the whole time horizon was re-run using different numbers of Basis splines for $\delta(t)$ and $\Omega(t)$. This served as a useful check on the sensitivity of the fit.
\item The optimization over the whole time horizon was re-run using a starting point based on the previously found optimum plus randomized deviations. This examines robustness of the final fit.
\end{enumerate}

\end{enumerate} 

\subsection{Error estimation}
To obtain error bars of the time-dependent functions we use non-parametric bootstrapping \cite{12Murphy}. The process is summarized as follow:
\begin{enumerate}
\item \textbf{Estimate initial solution} Estimate the time dependent functions from the original data using Hamiltonian estimation.
\item \textbf{Resampling} Create $N_{\rm s}$ sample solutions for all time-dependent functions in the following way:
\begin{enumerate}
\item Form a sample set by randomly picking with replacement from the photon count data used in qubit detection.
\item Re-estimate new time-dependent functions by optimizing over the full time span, using the solution found in (1) as a starting point.
\item Record the reduced chi-squared values $\chi_{\rm red,r}^2$ for each sample $r$ along with the B-spline curve coefficients $\vec{\alpha_{\rm r}}$
\end{enumerate}
\item \textbf{Post-process samples}
\begin{figure}[t!]
	\includegraphics[width = 1\columnwidth]{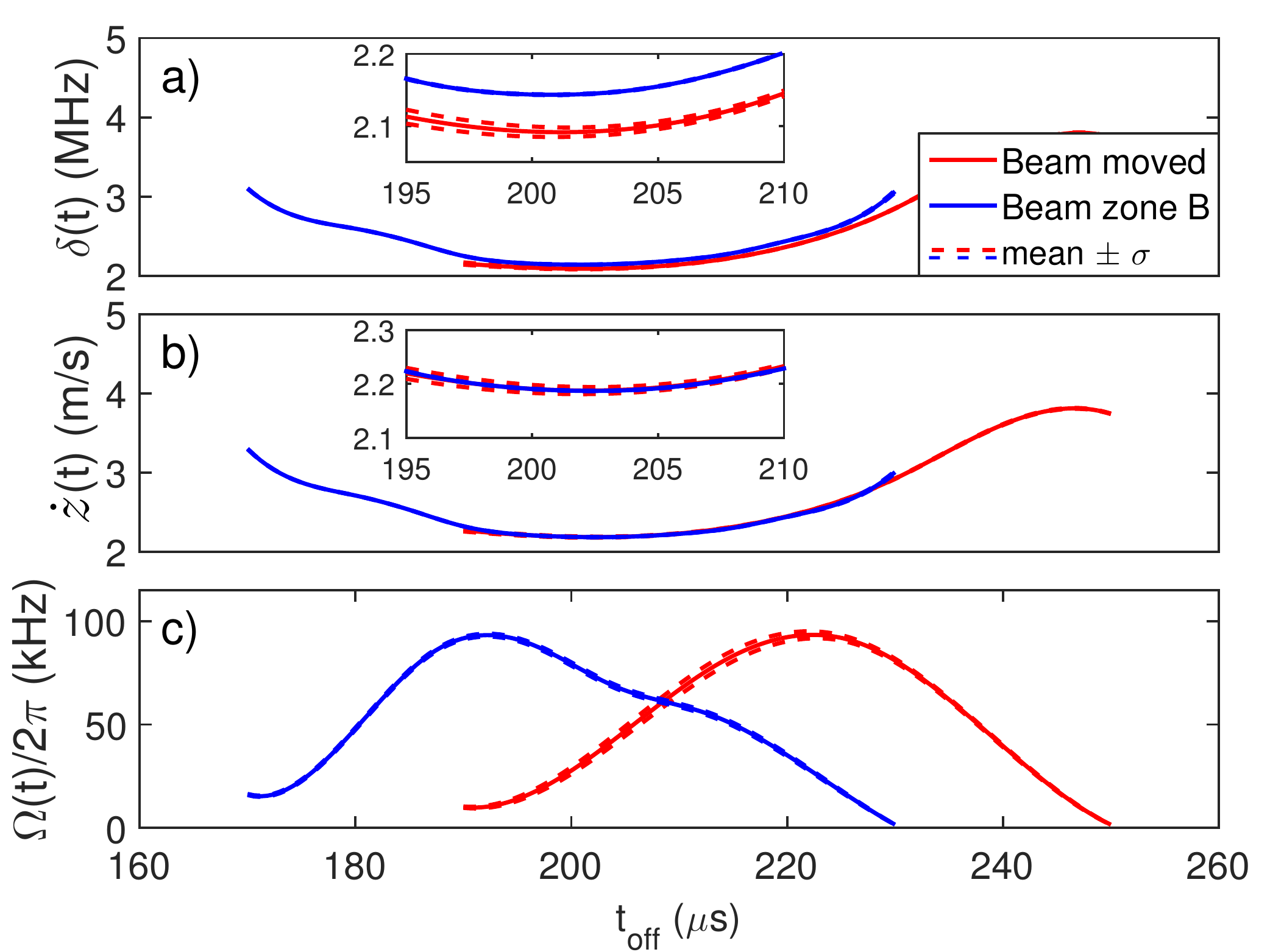}
	\caption{\textbf{Parametric bootstrap resampling:} Predictions for $\delta(t)$, $\dot{z}(t)$ and $\Omega(t)$ with error bounds obtained using parametric bootstrap resampling, assuming quantum projection noise. This can be compared with the error bounds obtained from the non-parametric method which are shown in Figure \ref{fig:before_after_beams} in the main text. The bounds are tighter for the parametric bootstrapping.}
	\label{fig:before_after_beams_NR}
\end{figure}
\begin{enumerate}
\item Form a histogram of the chi-squared values $\chi_{\rm red,r}^2$.
\item Find and fit a normal-like distribution to the histogram with preference to the spread with lowest lying $\chi_{\rm red,r}^2$ in the case of a multi-modal distribution. From the fit obtain the mean reduced-chi squared value $\langle\chi_{\rm red,r}^2\rangle$ as well as the standard deviation $\sigma_{\chi}$.
\item Eliminate the outlier samples by removing all $\vec{\alpha}_r$ with $\chi_{\rm red,r}^2$ values that are 3-5$\sigma_{\chi}$ from the mean $\langle\chi_{\rm red,r}^2\rangle$.
\item Form a matrix $\boldsymbol{Y}$ where each row vector is a sample set of coefficients $\vec{\alpha}_r$ that remained after step 3(c).
\end{enumerate}
\item \textbf{Obtain statistics}
\begin{enumerate}
\item Find the mean B-spline coefficients $\langle\vec{\alpha}\rangle$ of equation \ref{eq:spline_curve} by taking the mean over the column vectors of $\boldsymbol{Y}$ with each element of the mean given by $\langle\vec{\alpha}\rangle_{i}=\langle\alpha_i\rangle$.
\item Find the covariance matrix $\boldsymbol{\Sigma} = {\rm cov}(\boldsymbol{Y}^\alpha)$ with $\boldsymbol{\Sigma}_{ij}=\text{E}\left[\left(\alpha_i - \langle\alpha_i\rangle\right)\left(\alpha_j - \langle\alpha_j\rangle\right)\right]$ with E the expectation operator. The standard deviations of each of the mean coefficients  $\langle\alpha_i\rangle$ is given by $\sigma_{\langle\alpha_i\rangle} = \sqrt{\boldsymbol{\Sigma}_{ii}}$. We record these values in a row vector $\vec{\sigma}_{\langle\alpha_i\rangle}$.
\end{enumerate}
\end{enumerate}
\begin{figure*}
	\includegraphics[width = 1.95\columnwidth]{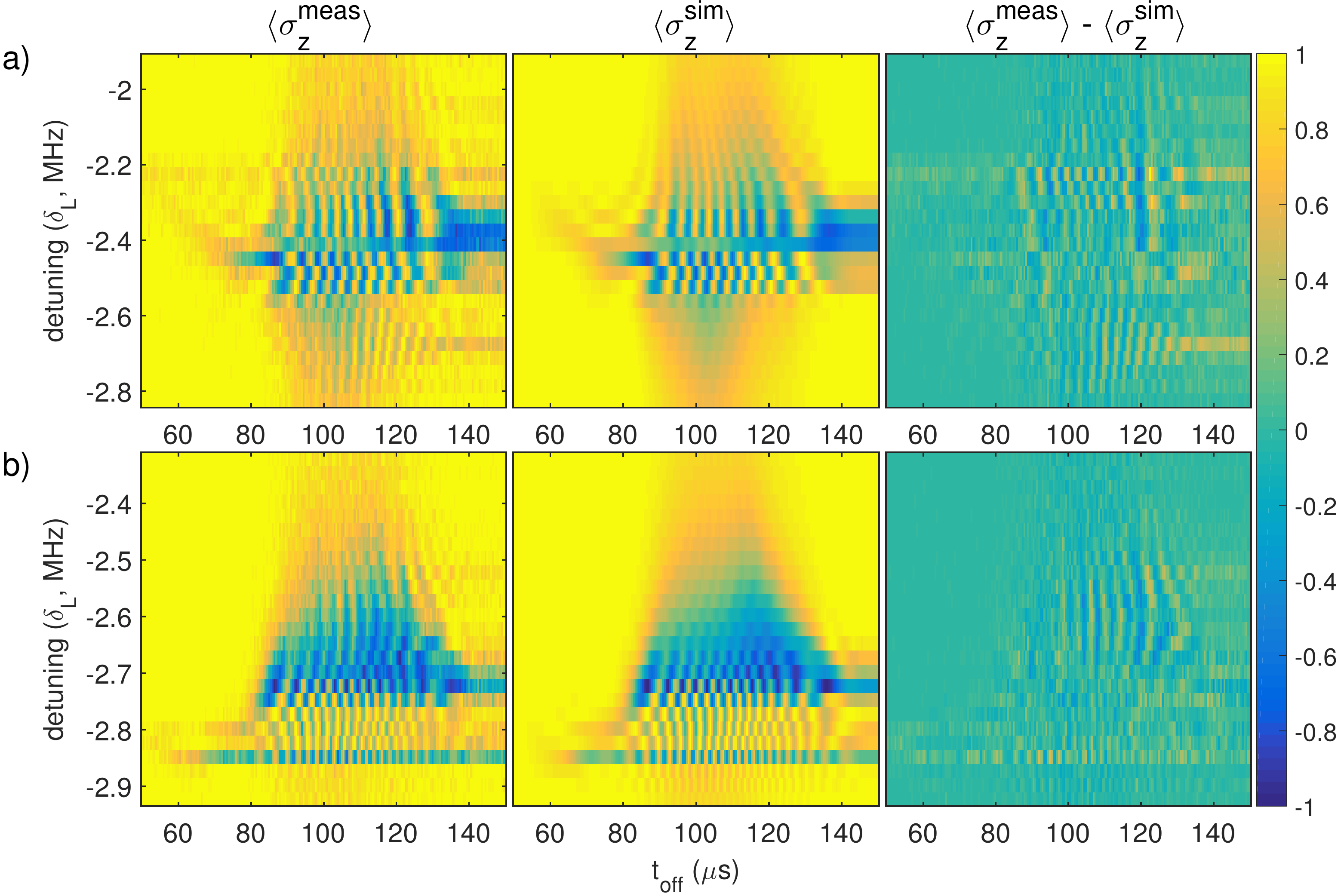}
	\caption{\textbf{Measured data, estimation and residuals:} Spin population as a function of detuning and switch-off time of the laser beam, for the data sets used to obtain the reconstructed parameters shown in figure \ref{fig:spatial_rabi_freqs}. a) uses a velocity profile with only small variations.  b) A second data run in which large variations in the velocity profile were used. Each data point results from 100 repetitions of the experimental sequence. For the Hamiltonian estimation the data was weighted according to quantum projection noise.}
	\label{fig:seconddata}
\end{figure*}

We have also applied parametric bootstrapping in order to obtain the error bounds shown in figure \ref{fig:before_after_beams_NR}. The difference to the non-parametric case is that in point (2) the samples are created using the solutions obtained from (1) and adding quantum projection noise. For each sample the Hamiltonian is estimated. The estimates from multiple samples are used to construct error bounds in the same manner as for the non-parametric resampling. We have found that the error bounds obtained from parametric bootstrapping are lower compared to that of the non-parametric case as shown in figure \ref{fig:before_after_beams}. We think this is due to the latter exploring deviations around a single minimum in the optimization landscape, while the case resampling arrives at different local minima which are spread over a wider region.

\subsection{Single beam profile with two different velocity profiles.}

As a check that our method is also able to produce consistent results for the Rabi frequency profile, we measured a second pair of data sets in which we take two different velocity profiles using the same beam position. This data is shown in Figure \ref{fig:seconddata}. Also shown are the best-fits obtained from the reconstructed Hamiltonians. The parameter variations obtained from the reconstructed Hamiltonians for these data sets can be found in the main text in Figure \ref{fig:spatial_rabi_freqs}. The sampling rate of the data in these data sets was 2~MHz, resulting in a Nyquist frequency of 1~MHz.

\end{document}